# Architecture of a Conference Management System Providing Advanced Paper Assignment Features

Yordan Kalmukov
University of Ruse,
8 Studentska Str., 7017 Ruse, Bulgaria

## ABSTRACT
This paper proposes an architecture and assignment management model of a conference management system that performs a precise and accurate automatic assignment of reviewers to papers. The system relies on taxonomy of keywords to describe papers and reviewers' competences. The implied hierarchical structure of the taxonomy provides important additional information – the semantic relationships between the separate keywords. It allows similarity measures to take into account not only the number of exactly matching keywords between a paper and a reviewer, but in case of non-matching ones to calculate how semantically close they are. Reviewers are allowed to bid on the papers they would like to (or not like to) review and to explicitly state conflicts of interest (CoI) with papers. An automatic CoI detection is checking for additional conflicts based on institutional affiliation, co-authorship (within the local database) and previous co-authorship in the past (within the major bibliographic indexes and digital libraries). The algorithm for automatic assignment takes into account all – selected keywords, reviewers' bids and conflicts of interest and tries to find the most accurate assignment while maintaining load balancing among reviewers.

## 1. INTRODUCTION
During the last several years conference management systems (CMS) became conference chairs' best friend. As a web based information systems they offer a reliable user-friendly service anywhere at any time. Authors can easily track the status of their papers. Program Committee (PC) members can review them anywhere in the world. But the highest benefit is for PC chairs as the conference management systems not just offer a user-friendly way of communication and data storage, but automation of a series of hard to handle and time consuming processes like assignment of reviewers to papers, conflict of interest detection, plagiarism detection and etc.

Conference management systems usually handle the entire process of conference management in smaller time-dependent pieces called phases. Some phases can overlap in time while others should be strictly sequential as they use data submitted or generated within previous phase(s). There are activities related to conference organizing (for example finding financial support; arranging halls for session presentations; arranging official dinners and other social activities; and etc.) that are more or less independent on the papers' management so they could be performed in parallel. At the time of assignment all papers have to be already submitted, all reviewers already registered and all bids stated so that the automatic assignment module proposes the best possible assignment taking into account both - selected topics and reviewers' bids. The reviewing and discussion phases can slightly overlap as some papers may already have all reviews done and waiting for discrepancy resolution while other papers may still be waiting for evaluation.

The process of assignment of reviewers to papers is probably the most important and challenging one. Its accuracy directly impacts the quality of the conference and its image. For high level conferences, having a low acceptance ratio, it is crucial that papers are evaluated by the most competent in the relevant subject domain reviewers. The assignment could be performed both manually and automatically. Manual assignment is feasible for conferences having a small amount of submitted papers. However when the number of papers and reviewers increases the manual assignment gets less and less accurate due to the constraints it should satisfy – high accuracy, no conflict of interest and not on the last place – load balancing (i.e. all reviewers should evaluate roughly the same number of papers).

The non-intersecting sets of papers and reviewers can be represented by a complete weighted bipartite graph (Figure 1), where P is the set of all submitted papers and R – the set of all registered reviewers. The weight of the edge between paper $p_i$ and reviewer $r_j$ indicates how competent (suitable) $r_j$ is to review $p_i$. This measure of suitability is also called a similarity factor. Similarity factors are usually calculated or assigned in accordance with the chosen method of describing papers and reviewers' competences [7].

Once the edges' weights are calculated the automatic assignment could be easily implemented by using assignment algorithms known from the graph theory – for example the Hungarian algorithm of Kuhn and Munkres [8, 9] or heuristic algorithms like the one proposed in [6].

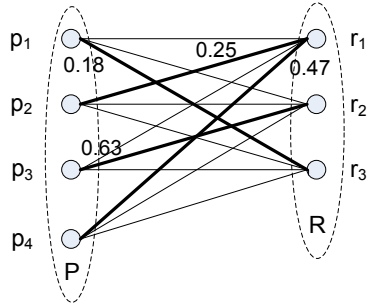

**Fig 1: The sets of papers (P) and reviewers (R) represented as a complete weighted bipartite graph. The edges in bold are the actual assignments suggested by an assignment algorithm. All edges have weights but just those of the assignments are shown for clearness.**

The accuracy of the automatic assignment depends on:
- *The method of describing papers and competences*. It determines whether or not an additional metadata is needed to describe papers and competences, and if users should explicitly provide it. It suggests the data model for representing objects descriptions and the similarity measures for calculating the similarity factors. The resulting similarity matrix shows how much exactly each reviewer is competent to evaluate each one of the papers.
- *The accuracy of the assignment algorithm itself*. The assignment algorithm takes the similarity matrix as an input and handles the assignment as an optimization problem [11] trying to find the best possible solution.

This article proposes *architecture* of a conference management system performing a precise and accurate automatic assignment of reviewers to papers, and a model for *effective management of the assignment process* as a whole. The system relies on explicit methods of describing papers and competences. It supports both bidding and topics (keywords) selection, but in contrast to all known similar solutions it describes the conference coverage area by *taxonomy of keywords* rather than just a non-structured list of topics. The implied hierarchical structure of the taxonomy provides important additional information – the semantic relationships between the separate keywords. It allows similarity measures to take into account not only the number of exactly matching keywords between a paper and a reviewer, but in case of non-matching ones to calculate how semantically close they are. *Thus if a paper and a reviewer do not share even a single keyword in common a non-zero similarity factor could still be accurately calculated.* Taxonomy allows the conference to be described with many more semantically-related (even semantically overlapping) keywords. The increased number of keywords results in more detailed keywords, i.e. more precise description of both papers and competences.

The present paper focuses just on the architecture of the conference management system and the management of the assignment process as a whole. Another article will be entirely devoted to the proposed method of describing papers and competences by taxonomy of keywords. It will: justify the method's advantages in comparison to other methods of describing used in the existing conference management systems; propose similarity measures for precise calculation of paper-reviewer similarity factors; perform a complexity and accuracy evaluation of the suggested method.

## 2. RELATED WORK

About a dozen conference management systems have been developed during the last years – The MyReview System [38], OpenConf [33], IAPR Commence [28], EasyChair [25], CyberChair [15, 22], EDAS [26], Confious [10, 18], ConfSys [4, 5], ConfTool [20], ConfMaster [19] and Microsoft's CMT [32].

In accordance with the current trend in computer software, most of them are offered not as a traditional sellable software product but as a completely hosted service. The *Software as a Service (SaaS)* solutions minimize the total cost of service by releasing conference organizers from buying any hardware and software, hiring system administrators and programmers to install and support them and etc. However other important issues arise in respect to these cloud based solutions – *security* and *privacy* discussed by Mark Ryan in [14]. For organizers who prefer to host the conference management system on their own servers and to take full responsibility for their data The MyReview System and OpenConf are invaluable products.

Beside the differences in graphics and user interface all conference management systems provide all the basic functionalities needed to manage the processes of paper submission, assignment of reviewers to papers, review submission and etc. Most systems offer an automatic pre-press processing of papers and conference proceedings, automatic generation of conference program (based on manual paper clustering), authors' index and table of contents. CMT and EDAS check submitted papers for plagiarism by utilizing external services like iThenticate [30] and Docoloc [24].

The most significant difference between all mentioned conference management applications however is in the way they handle the automatic assignment of reviewers to papers, i.e. in the methods of describing papers and competences they use and the assignment algorithms they employ.

Generally the methods of describing could be divided into two main groups:
- Explicit methods – require users to explicitly outline their papers and/or competences, i.e. to provide some descriptive metadata;
- Implicit methods – intelligent methods that automatically extract the required descriptive data from the papers and from reviewers' previous publications available on the Internet.

The implicit methods usually perform a text analysis of papers' content and/or the content of online digital libraries, bibliographic indexes and others resources - DBLP [23], ACM

Digital Library [16], CiteSeer [36], Google Scholar [27], Ceur WS [17] and etc.

Andreas Pesenhofer et al. [11] suggest that *the interest of a reviewer can be identified based on his/her previous publications available on the Internet*. Stefano Ferilli et al. [3] prpose that *paper topics could be extracted from its title and abstract, and expertise of the reviewers from the titles of their publications available on the Internet*. Marko Rodriguez and Johan Bollen [13] suggest that *a manuscript's subject domain can be represented by the authors of its references*.

Implicit methods rely on external data sources on the Internet that are more or less inertial and contain sparse information. New papers and articles are indexed with months delay and not by one and the same bibliographic index. The proposers of the implicit methods admit they were unable to find relevant information on the Internet for about 20% of the PC members of the conferences implementing those methods. Probably for that reason all commercially available conference management systems rely on explicit methods only. However implicit methods could be very useful for automatic detection of conflicts of interest.

Existing systems use the following three different ways of explicitly describing papers and reviewers' competences:

- Bidding / rating papers (reviewers state their interest to each paper individually);
- Choosing from a predefined list of conference topics;
- Combined, conference topics + bidding.

Bidding requires reviewers to browse all submitted papers and to state their interest to review (or not) each one of them individually. This is usually done by selecting the relevant bid option from a drop down menu. Reviewers' bids are considered to be absolutely accurate as nobody knows better than the reviewer himself if he is competent to review a specified paper or not. Bidding however does not really describe papers nor reviewers, but just explicitly state the relationship between them. If a paper has not been rated by enough reviewers then they will be assigned to it at random as there is no way the conference management system to determine the paper's subject domain or reviewers' competences. In case the number of submitted papers is high reviewers never bid on all of them that results in plenty of missing similarity factors. Philippe Rigaux proposes a clever improvement, called an Iterative Rating Method (IRM) [12], of the bidding process which is trying to overcome the random assignments by predicting the missing bids by iteratively applying a collaborative filtering algorithm to the ratings that has been explicitly provided by reviewers or those that have been previously predicted (within previous iterations).

Describing papers and reviewers' competences by a list of conference topics (keywords) assumes that PC chairs set up a predefined list (used as a binary set) of topics that best describe the conference coverage area. During paper submission authors are required to select (by using HTML checkboxes) those topics that best outline their papers. During registration reviewers should select the topics corresponding to the areas of science they are competent in. In contrast to bidding, describing by a list of topics provides an independent, stand-alone description of every single paper and reviewer. Defining an appropriate and good enough list of topics however is a challenging task. Keywords should not be too many, but they have to cover the entire area of science covered by the conference, and they should provide enough details for precise description of papers and competences. For broad-area multi-disciplinary conferences that is mission impossible.

The combination of both topics matching and bidding aims to bring the advantages of the two methods together. Topics matching should be used first to identify a small subset of papers to be suggested for bidding to each reviewer. If bid is available it explicitly determines the similarity factor between the specified paper and reviewer. If the reviewer has not bided on the given paper then the similarity factor between them is calculated by using topics matching only.

The MyReview System, OpenConf Professional Edition, EasyChair, CyberChair, EDAS, Confoious, ConfSys, ConfTool and Microsoft's CMT all rely on both list of conference topics and reviewers' bids, however there are significant differences in the way they handle the assignment.

Based on topics matching the **MyReview** system selects a subset of papers to be explicitly rated by each reviewer. Then missing bids are either predicted by the IRM or implicitly determined by topics matching (papers are described just by a single topic while competences by multiple topics). If the paper's topic matches one of the reviewer's topics the bid is set to "Interested". If there is no topic in common the bid is set to "Why not" [12, 38]. The latter seems not to be a very relevant decision as the "why not" bid still indicates minor competences in the paper's subject domain that could not be justified as there is no explicit bid and no common topic. Finally the maximum-weighted matching algorithm of Kuhn and Munkres [8, 9, 12] is applied over the bids to propose the best possible assignment.

Similarly to the MyReview system, the **EasyChair**'s assignment algorithm employs bids only. If those are missing they are implicitly determined by topics matching as follows: if a paper has more than one common topic with the PC member, it will be regarded as if he chooses "I want to review this paper"; if a paper has exactly one common topic with the PC member, it will be regarded as if he chooses "I can review it" [25].

**EDAS** and **CyberChair** use custom-designed greedy algorithms that process papers in turn, starting with the one with the smallest number of bids. The algorithms first assign reviewers based on bids only (including low-level preferences) then if the number of assigned reviewers is not enough the process continues with topics matching. If two or more reviewers have claimed the same bid level (willing to review) the one with the smallest number of papers is assigned first [22, 26]. The similarity measure based on topics matching is not specified.

**OpenConf** Professional Edition, **Confious**, **ConfSys**, **ConfTool** and **CMT** do not specify any details on how they use topics matching and bidding together in the automatic assignment of reviewers to papers. Rigaux reports that CMT uses a greedy algorithm, assigning a paper to the reviewers who gave the higher preference, but limiting the number of papers assigned to a reviewer to a threshold. When the system cannot find a reviewer, a matching of both the reviewers and paper topics is used. If this fails the result is unpredictable [12].

**OpenConf Community Edition** relies on list of topics only together with a custom similarity measure to calculate weights between papers and reviewers. Then the assignment is handled by a custom greedy algorithm that first assign reviewers to those papers described by the smallest number of topics or to papers described by topics chosen by a small amount of reviewers.

**IAPR Commence** offers bidding together with manual assignment only.

All conference management systems allow reviewers to explicitly state conflict of interest (CoI) with papers during the bidding phase. Additionally an automatic CoI detection is also provided based on co-authorship (within the local database) and/or institutional affiliation. Confious and ConfSys check for previous co-authorship by fetching authors' and reviewers' publications available in DBLP. Both explicitly stated or automatically detected conflicts of interest prevent the assignment algorithms from matching the specified papers and reviewers together.

An in-depth comparative analysis of the methods of describing papers and competences used by the existing conference management systems could be found in [7].

## 3. AN ARCHITECTURE OF A CONFERENCE MANAGEMENT SYSTEM

The architecture of the proposed conference management system is shown on figure 2. At a glance it is a standard 3-tier architecture similar to those of many other web-based information systems. The business (domain-specific) logic is separated from both the user interface and the data. The conference management system stores its vital data in a local database but uses information from external web services as well. It implements an additional layer of data abstraction that makes it independent on the database management system (DBMS) and the specific data access language it uses.

The *User Management & Access Control* is a domain-independent *subsystem* that handles user registration; authentication, authorization and access control; profile management, user rights management and etc. It could be easily ported (without any changes) to every other system in any subject domain. This of course applies to all other domain-independent components – *Database Abstraction Layer / Data Access Object*; *Graphic User Interface*; and *XML Parser / Composer*.

The *Reviewers' Functionalities* component delivers all the basic functionalities that reviewers need to do their job – bidding on papers, reviewing papers and etc.

As the name of the *Papers Management* component suggests it provides management utilities like – browsing / searching / sorting papers; changing papers' statuses (accepted, rejected, in discussion) both manually and automatically; editing and deleting papers; and not on the last place *grouping papers in sessions* and conference programme generation. Grouping could be done by performing a cluster analysis utilizing the very same similarity measure as the one used to calculate similarity factors between papers and reviewers, but applied on paper-to-paper basis. Another approach is by using reviewers' bids and a collaborative filtering algorithm. A combination of both ways is feasible as well.

The *Plagiarism Detection* component extracts papers' content, converts it in the relevant format and sends it to external third-party web services like CrossRef's CrossCheck [21], iThenticate [30], PlagScan [35], PlagAware [34], Docoloc [24] or others for extensive check for plagiarism. This component does not take any decisions automatically but just shows the returned results to the conference chairs so they could perform further investigation if necessary or reject the plagiarized papers.

The *Discrepancies Resolution & Virtual PC Meeting* is used during the discussion phase. Its main idea is to allow PC members to read the reviews of each other and to discuss which papers should be accepted or rejected. All reviews and discussions are completely anonymous. Real names are visible to the PC chairs only. This component could be set up to work in various different ways within different timeframes. For example, at the beginning just the reviewers of a same paper could be allowed to read the evaluation of each other and to discuss if the paper should be accepted or not. If they can not resolve the conflict of their reviews and make a decision then other reviewers could be added to the discussion.

As mention in the introduction, the proposed conference management system describes the conference coverage area by taxonomy of keywords. It allows similarity measures to take into account not only the number of exactly matching keywords between a paper and a reviewer, but in case of non-matching ones to calculate how semantically close they are. The *Local Taxonomy Management* provides all the necessary management utilities for building the desired conference taxonomy right through the system's user interface or importing and editing pre-defined taxonomies from external XML files.

The *Paper Assignment* component, as the most intriguing one, is reviewed in details. Its internal structure is shown on figure 3.

The *Keywords Analysis and Manipulation*, as its name suggests, performs an automatic analysis of the keyword sets chosen by both authors and reviewers - an analysis that often results in modifying the original sets by adding or removing keywords to/from them.

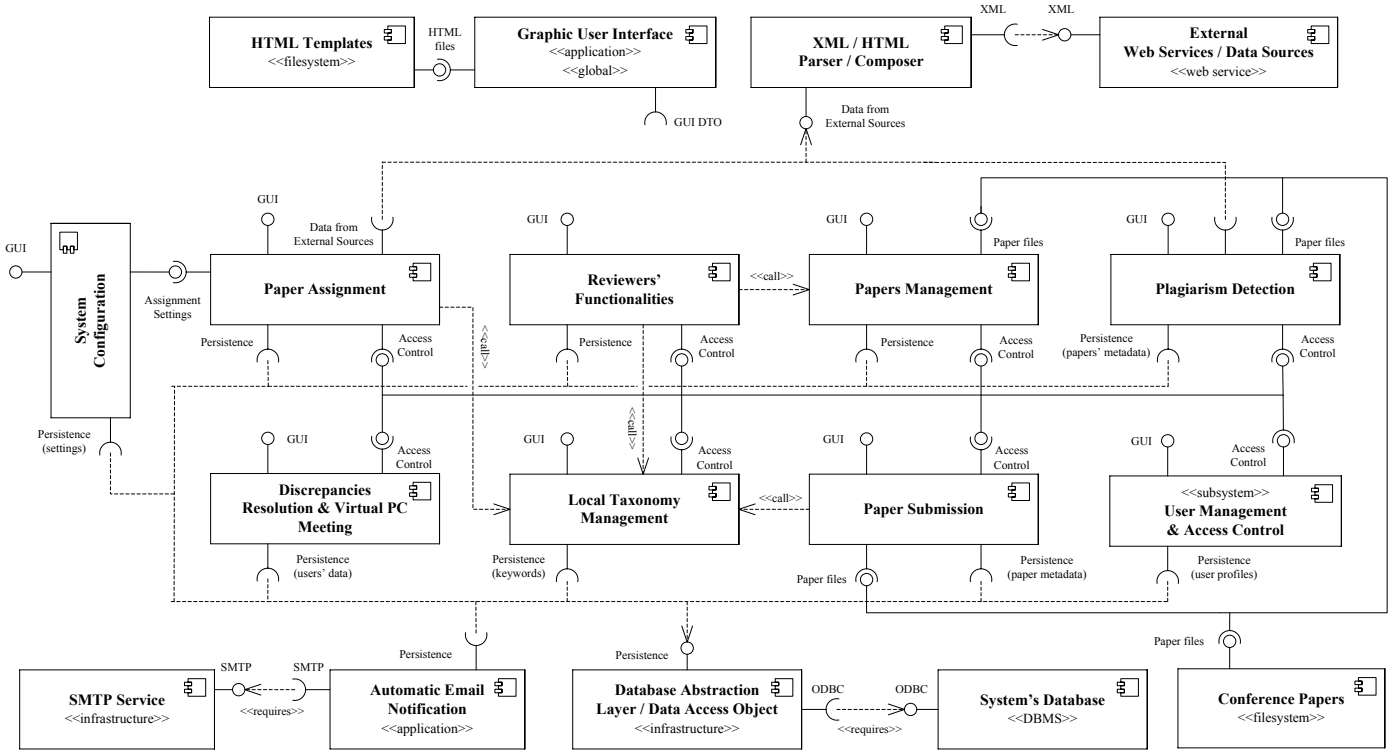

**Fig 2: Architecture of the conference management system**

An important issue arises immediately – if such actions of manipulating keywords are justified (or even legal) without any formal approval by authors and reviewers. They are justified as long as they do not maliciously substitute users' point of view but just refine it. In fact this entire sub-module is completely inspired by users' behavior. Original sets are modified only if they satisfy a predefined set of rules. Adding keywords is typically applied on the sets chosen by reviewers only. Why? Because reviewers tend to generalize their competences but achieving higher accuracy requires users to select nodes deeper in the hierarchy as they are more informative. If, for example, the reviewer feels highly competent on all aspects of information systems then he/she is more likely to choose the general node "Information Systems" rather than its sub-nodes. Thus the following *modification rule* does not fake reviewers' preferences but just help increasing accuracy:

*If*

    1. A reviewer has stated high or medium level of competence for a node $n_i$

and 2. $n_i$ is less than […] levels in depth from the root

and 3. $n_i$ has children

and 4. the reviewer has **not** selected any of $n_i$'s children.

*then*:

    Add $n_i$'s direct successors to the set of keywords chosen by the reviewer (assuming the reviewer is generalizing his knowledge and instead of selecting all children he has selected just their common ancestor for convenience)

Removing keywords is only done if it is required by the similarity measures. Reducing is necessary to handle situations where for example an author has selected a single node deep in the hierarchy while the reviewer has selected the same node (or closed to it) but also all nodes in the path from the root to the specified one. In this case the increased number of selected keywords by the reviewer will incorrectly decrease the overall similarity factor. The set describing a paper may be derived from its original in a way that no keyword is in a direct parent-child relationship with another keyword by removing the parent as the child is more informative. Then the reviewer's set may contain just those keywords which are the semantically closest pairs of the keywords in the paper's set. Alternative ways for reducing sets may be also applied.

The **Bidding / Explicit CoI Processor** serves as a translator that converts bids to similarity factors so that they replace the corresponding automatically calculated similarities. As bids directly reflect reviewers' preferences they are considered to be more accurate than any automatically calculated similarities thus if a reviewer has explicitly bided on a paper then the bid completely determines the similarity factor between these paper and reviewer. There are two general ways of converting bids to similarities – static mapping and dynamic calculation. Static says that every bid "Expert, willing to review" should be translated as 1.0; "Expert" as 0.9; "Not expert but capable" as 0.6 and so on. Dynamic on the other hand first analyzes all bids in respect to a single paper and then calculate the exact similarity factor corresponding to the bid.

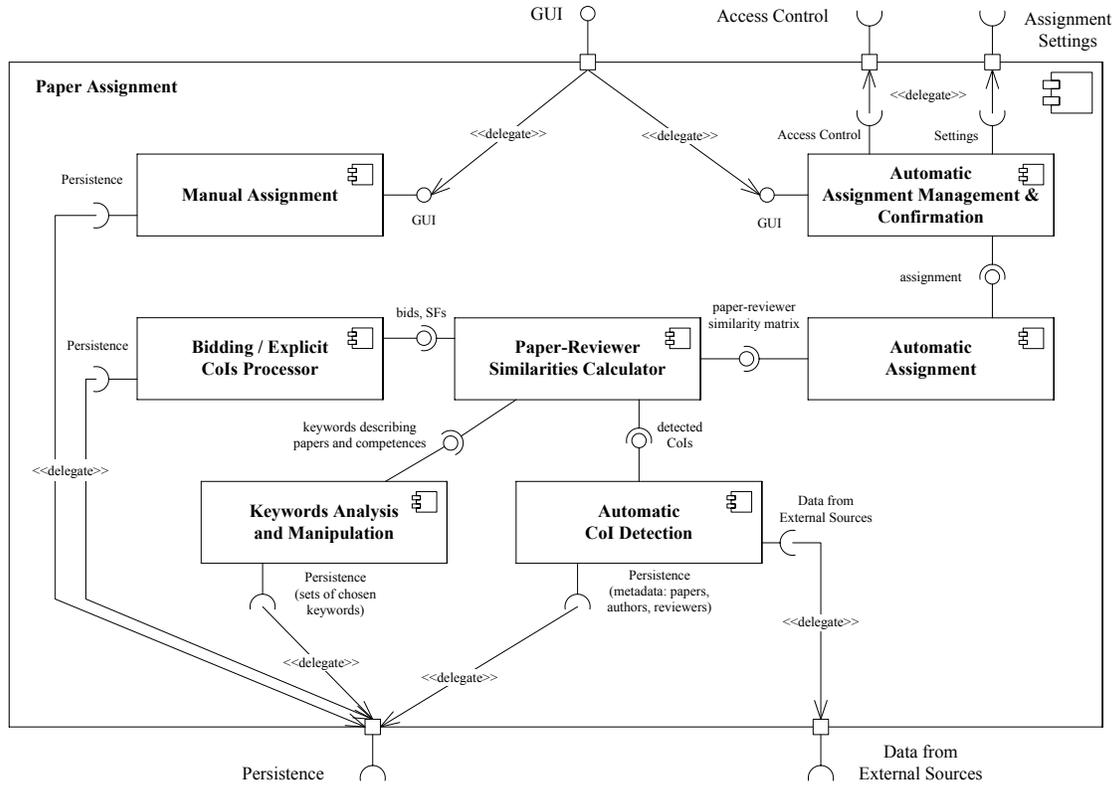

**Fig 3: Internal structure of the Paper Assignment component**

Expectedly an explicitly stated conflict of interest is converted to a zero similarity factor so that reviewer is never assigned to the given paper.

The *Automatic CoI Detection* looks for conflicts of interest of the following types:

- *Same country* (if applicable). Some conferences strictly require that author and reviewer should be from different countries.
- *Same institution*. Usually by performing a text analysis over the names of the author's and reviewer's institutions and/or by comparing the domain-name part of user email addresses.
- *Co-authorship* (within the local database). Checks if the reviewer is a co-author of the paper, or is a co-author of a co-author of the paper.
- *Previous co-authorship into the past*. Similar to the previous one but in respect to a much longer period of time and most importantly by performing a global check within the major bibliographic indexes. That guarantees if a co-authorship occurs in another conference or journal in the past it will be detected as well.

Global CoI detection could be achieved by extracting data from social networks as FOAF [37], LinkedIn [31] and etc. and/or large bibliographic indexes like DBLP and Google Scholar. Aleman-Meza et al propose a good approach for CoI detection by applying semantic analytics techniques on social networks [1, 2].

The *Paper-Reviewer Similarities Calculator* implements the similarity measures and all other mathematics needed for precise calculation of similarity factors from the sets of chosen keywords, reviewers' bids and detected / stated conflicts of interest (all merged together). The resulting *similarity matrix* shows how much each reviewer is competent (suitable) to evaluate each one of the submitted papers. The matrix is the only input to the *Automatic Assignment* component that handles the assignment as a global optimization task by applying a maximum-weighted matching algorithm like the Hungarian algorithm of Kuhn and Munkres [8, 9] or by faster heuristic algorithm like the one proposed in [6]. The assignment algorithm should be run on several passes so that every paper is assigned to the desired number of reviewers. The module proposes an assignment but does not apply it without the formal approval of a PC chair. Approval could be complete (for the entire proposal) or partial (for specific edges only).

The process of automatic assignment is controlled by the *Automatic Assignment Management & Confirmation* component. After approving the assignment PC chairs are allowed to further modify it and reassign papers and reviewers if needed. The *Manual Assignment* provides all the needed functionalities for flexible manual assignment management.

The assignment management process is outlined in details on figure 4.

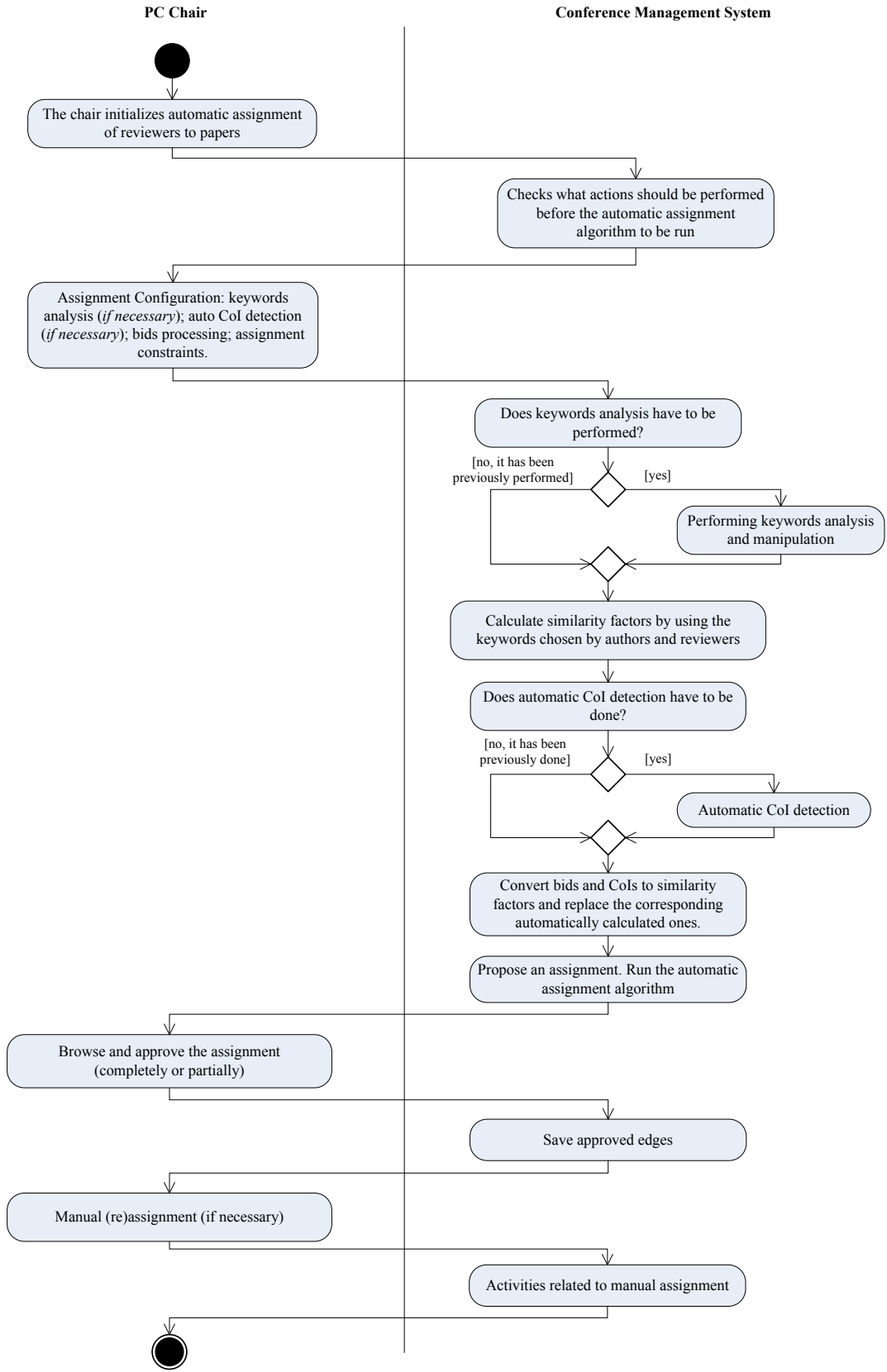

**Fig 4: The assignment management process**

## 4. CONCLUSIONS

The proposed architecture components (except the automatic plagiarism detection), assignment management model and method of describing papers and competences by taxonomy of keywords are fully implemented and used by CompSysTech [29], a broad-area, multi-disciplinary, ICT-related conference in year 2010 and 2011. In respect to the automatic paper assignment the conference management system achieved very good results. During review submission PC members were required to evaluate their expertise in the papers they were assigned to review. 88% of the reviewers stated high or medium level of expertise. The rest 12% specified low level but that is unavoidable due to the uneven distribution of papers and competences over the conference topics (especially true for broad area conferences like CompSysTech). Furthermore about 83% of the automatically calculated similarity factors tightly correspond to the levels of expertise explicitly stated by reviewers. A brief analysis of the results shows that the primary reason for inaccurate calculation of some similarity factors is the selection of too general taxonomy nodes like "Software" or "Computer Applications". These nodes do not provide enough details resulting in less precise assignment. The accuracy could be further increased if selection of too general nodes is technically disallowed.

Additional analysis and observation of the users' behavior in future could significantly help to improve the *Keywords Analysis and Manipulation* module so it can compensate even more mistakes in keywords selections caused by subjective reasons. Decreasing the influence of the latter will further increase the assignment accuracy then subsequently the conference image and the authors' thrust in it.

## 5. ACKNOWLEDGMENTS

This paper is financed by project: Creative Development Support of Doctoral Students, Post-Doctoral and Young Researches in the Field of Computer Science, BG 051PO001-3.3.04/13, European Social Fund 2007–2013, Operational Programme "Human Resources Development".